\documentclass[prl,twocolumn,showpacs,preprintnumbers,aps,nofootinbib]{revtex4-1}
\usepackage{graphicx}
\usepackage{dcolumn}
\usepackage{bm}
\usepackage{amsmath,amssymb,amsfonts}
\usepackage{latexsym}
\usepackage{bbm}
\usepackage{color}

\begin{document}

\preprint{ACFI-T17-12} 

\title{Electroweak Baryogenesis driven by Extra Top Yukawa Couplings}

\author{Kaori Fuyuto$^1$}
\email{kfuyuto@umass.edu}
\author{Wei-Shu Hou$^2$}
\email{wshou@phys.ntu.edu.tw}
\author{Eibun Senaha$^3$}
\email{senaha@ibs.re.kr}
\affiliation{$^1$Amherst Center for Fundamental Interactions, Department of Physics,
University of Massachusetts Amherst, MA 01003, USA}
\affiliation{$^2$Department of Physics, National Taiwan University, Taipei 10617, Taiwan}
\affiliation{$^3$Center for Theoretical Physics of the Universe, Institute for Basic Science (IBS), Daejeon 34051, Korea}
\bigskip

\date{\today}

\begin{abstract}
We study electroweak baryogenesis driven by the top quark in a general two Higgs doublet
model with flavor-changing Yukawa couplings, keeping the Higgs potential $CP$ invariant.
With Higgs sector couplings and the additional top Yukawa coupling $\rho_{tt}$
all of $\mathcal{O}$(1), one naturally has sizable $CP$ violation
that fuels the cosmic baryon asymmetry.
Even if $\rho_{tt}$ vanishes, the favor-changing coupling $\rho_{tc}$
can still lead to successful baryogenesis.
Phenomenological consequences such as $t\to ch$, $\tau \to \mu\gamma$
electron electric dipole moment, $h\to\gamma\gamma$,
and $hhh$ coupling are discussed.
\end{abstract}


\maketitle

\paragraph{Introduction.---}
The discovery of a scalar particle 125~GeV in mass~\cite{h125_discovery}
is a first step towards the thorough understanding of
spontaneous electroweak symmetry breaking (EWSB).
Current data suggest~\cite{Khachatryan:2016vau} the observed scalar
belongs to an $\text{SU}(2)_L$ doublet that is responsible for
EWSB and particle mass generation.
Understanding the full structure of the Higgs sector is a
primary goal of particle physics and cosmology.

Even though one Higgs doublet alone is sufficient to play
the two aforementioned roles,
it is natural to consider a multi-Higgs sector, since the
Standard Model (SM) itself has serious drawbacks.
Two such drawbacks are insufficiency of $CP$ violation (CPV)
and lack of out of equilibrium process, such that
the baryon asymmetry of the Universe (BAU) cannot arise.
It is known that these two shortcomings can be resolved
if the number of Higgs doublets is at least two,
and one can have~\cite{ewbg} {\it electroweak baryogenesis} (EWBG),
with the attraction of sub-TeV dynamics that can be tested at the LHC.

In a two Higgs doublet model (2HDM), the electroweak phase transition (EWPT)
can be first order~\cite{EWPT_2HDM},
inducing departure from
equilibrium around Higgs bubble walls that separate symmetric from broken phases.
In this Letter, we advocate the absence of
\emph{ad~hoc} discrete symmetries~\cite{Glashow:1976nt}.
With both doublets coupling to fermions,
there are extra complex Yukawa couplings
that yield CPV beyond the Cabibbo-Kobayashi-Maskawa (CKM) framework.
Besides providing new CPV sources,
the extra off-diagonal elements are in general nonzero,
giving rise to flavor changing neutral Higgs (FCNH)
processes such as $t\to ch$~\cite{Hou:1991un}.
Such FCNH couplings can accommodate
many experimental anomalies~\cite{Fajfer:2012jt, Crivellin:2012ye, hmutau_2HDM}.

In this Letter, we study EWBG in this general 2HDM,
focusing on up-type heavy quarks.
The CPV source terms that fuel BAU are
estimated using a closed time path formalism with
vacuum expectation value (VEV) insertion approximation.
Depending on up-type Yukawa textures with $\mathcal{O}(1)$ complex couplings,
CPV relevant to BAU is efficiently sourced by top-charm flavor
changing transport,
which is in stark contrast to a 2HDM with $Z_2$ symmetry
that forbids such couplings and phases, and
CPV has to arise from the Higgs sector.

We explore the parameter space of Yukawa structures that favor EWBG
and discuss phenomenological consequences such as $t\to ch$,
electron electric dipole moment,
$h \to \gamma\gamma$ and $hhh$ coupling.
We also compare with the scenario~\cite{Chiang:2016vgf}
motivated by a hint for $h\to\mu\tau$~\cite{hmutau_LHC}
which has since disappeared~\cite{CMS:2017onh},
and discuss $\tau\to \mu\gamma$.

\paragraph{Model.---}
Without imposing any $Z_2$ symmetry,
the fermions can couple to both Higgs doublets,
and the Yukawa interaction for up-type quarks is
\begin{align}
-\mathcal{L}_Y
& = \bar{q}_{iL}(Y^u_{1ij}\tilde{\Phi}_1+Y^u_{2ij}\tilde{\Phi}_2 ) u_{jR}+{\rm h.c.},
\label{Yukawa}
\end{align}
where $i, j$ are flavor indices, $\tilde \Phi_b = i\tau_2\Phi^*_b$ ($b=1, 2$) with
\begin{align}
\Phi_{b}(x)=
\left(\begin{array}{c}
\phi^+_b(x)\\
\frac{1}{\sqrt{2}}\left(v_b+h_b(x)+ia_b(x)\right)
\end{array}\right),
\end{align}
and $\tau_2$ is a Pauli matrix.
Denoting the VEVs as
$v_1=v\, c_\beta$ and $v_2=v\, s_\beta$ ($v\cong 246~$GeV),
hereafter we use the shorthand
$s_\beta=\sin\beta$, $c_\beta=\cos\beta$ and $t_\beta=\tan\beta$.

In the basis where only one Higgs doublet has VEV,
the $CP$-even Higgs fields $h^{\prime}_{1,\,2}$ are related to
the mass eigenstates
through a mixing angle $\beta-\alpha$:
$h'_1= c_{\beta-\alpha}\, H + s_{\beta-\alpha}\, h$ and
$h'_2=-s_{\beta-\alpha}\, H + c_{\beta-\alpha}\, h$,
where $h$ is the observed 125 GeV scalar.
From Eq.~(\ref{Yukawa}), we have
\begin{align}
V_L^{u\dagger}Y^{\text{SM}}V^u_R={\rm diag}(y_u,~y_c,~y_t)\equiv Y_{\rm diag},
\label{YD}
\end{align}
where $Y^{\text{SM}}= Y_1\, c_{\beta}+Y_2\, s_{\beta}$
is diagonalized by a biunitary transform
to give quark masses $m_f=y_f\, v/\sqrt{2}$.
The neutral up-type Yukawa interaction becomes
\begin{align}
-\mathcal{L}_Y
&
 = \bar{u}_{iL}
\left[\frac{y_i\delta_{ij}}{\sqrt{2}} s_{\beta-\alpha}
+\frac{\rho_{ij}}{\sqrt{2}}\, c_{\beta-\alpha}
\right] u_{jR}h \nonumber \\
&\quad +\bar{u}_{iL}
\left[\frac{y_i\delta_{ij}}{\sqrt{2}}\, c_{\beta-\alpha}
-\frac{\rho_{ij}}{\sqrt{2}}\, s_{\beta-\alpha}
\right] u_{jR}H \nonumber \\
&\quad -\frac{i}{\sqrt{2}}\, \bar{u}_{iL}\rho_{ij} u_{jR}\, A+{\rm h.c.}, \label{Yukawa_low}
\end{align}
where
\begin{align}
\rho = V_L^{u\dagger}\left(-Y_1\, s_{\beta}+Y_2\, c_{\beta} \right)V^u_R,
\label{rhoij}
\end{align}
is in general flavor changing,
and we parameterize $\rho_{ij}=|\rho_{ij}|e^{i\phi_{ij}}$.
In the ``alignment'' limit of $c_{\beta-\alpha} \to 0$,
$h$ becomes the SM Higgs boson,
and all FCNHs are relegated to the heavy Higgs sector.
It has been shown~\cite{Hou:2017hiw} recently that
alignment is a natural consequence of the
general 2HDM with similar parameter settings.

With no $Z_2$ symmetry,
the Higgs potential takes the general form.
Current LHC data indicate that the observed boson $h$ is $CP$-even~\cite{h125SP}.
Moreover, CPV phases in the Higgs potential are
highly constrained by EDMs of electron, neutron, etc.~\cite{HiggsCPV}.
We therefore assume a $CP$ conserving Higgs sector for simplicity. 
Down-type Yukawa interactions can also hold FCNH couplings 
analogous to Eq.~(\ref{rhoij}). However, the down sector 
receives much stronger constraints from $B$ physics, 
such as $B_s-\bar{B}_s$ mixing and $b\to s \gamma$ transition. 
Thus, we expect the production of our present BAU to be less efficient
from down sector, and our study focuses exclusively on 
extra up-type Yukawa couplings.

\paragraph{Electroweak baryogenesis.---}
BAU is generated by a sphaleron process in the symmetric phase,
where the VEVs are zero.
To avoid washout, similar processes have to be suppressed in the broken phase.
A rough criterion is given by the condition $\Gamma_B^{(\text{br})}(T_C)<H(T_C)$,
i.e. the baryon number changing rate $\Gamma_B^{(\text{br})}(T_C)$ is less than
the Hubble parameter $H(T_C)$ at critical temperature $T_C$.
This condition can be satisfied if the EWPT is first order such that
$v_C/T_C \gtrsim 1$
where $v_C=(v_1^2(T_C)+v_2^2(T_C))^{1/2}$.
Thermal loops of heavy Higgs bosons can make the
first-order EWPT strong enough~\cite{EWPT_2HDM} to satisfy this criterion,
owing to ${\cal O}(1)$ nondecoupled~\cite{Kanemura:2004ch} Higgs couplings.
Such couplings would lead to intriguing phenomenological consequences,
such as variation of $h \to \gamma\gamma$ width ($\mu_{\gamma\gamma}$)~\cite{Ginzburg:2002wt},
and triple Higgs boson coupling ($\lambda_{hhh}$)~\cite{Kanemura:2004ch}
compared with SM values, as we will quantify below.

Departure from equilibrium is in the form of an expanding
bubble of the broken phase due to first order EWPT.
We estimate BAU by (see e.g. Refs.~\cite{Huet:1995sh, Cline:2000kb})
\begin{align}
Y_B \equiv \frac{n_B}{s} = \frac{-3\Gamma_B^{(\text{sym})}}{2D_q\lambda_+s}
\int_{-\infty}^0dz'~n_L(z')e^{-\lambda_-z'},
\label{YB}
\end{align}
where $D_q \simeq 8.9/T$ is the quark diffusion constant, $s$ is the entropy density, $\Gamma_B^{(\text{sym})} = 120\alpha_W^5T$ is the $B$-changing rate in the symmetric phase
and $\lambda_\pm = \big[v_w \pm (v_w^2 + 15\Gamma_B^{(\text{sym})}D_q)^{1/2}\big]/2D_q$,
with $\alpha_W$ the weak coupling constant and $v_w$ the bubble wall velocity.
The integration is over $z'$, the coordinate opposite the bubble expansion direction,
and nonvanishing total left-handed fermion number density $n_L$ is needed for $Y_B$.
We use the Planck value $Y_B^{\rm obs}= 8.59\times 10^{-11}$~\cite{Ade:2013zuv}
for our numerical analysis of viable parameter space for EWBG.

\begin{figure}[t]
\center
\includegraphics[width=5cm]{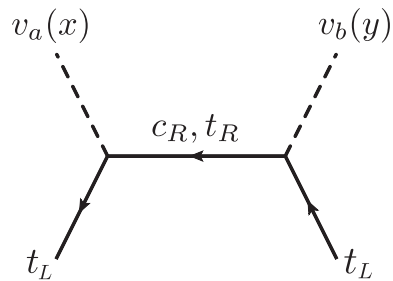}
\caption{
A dominant CPV process relevant for the baryon asymmetry, with
Higgs bubble wall denoted symbolically as $v_{a}(x)$ and $v_b(y)$.
The vertices can be read off from Eq.~(\ref{Yukawa}).
}
\label{fig:bubble_int}
\end{figure}

The BAU-related CPV arises from the interaction between
particles/antiparticles and the bubble wall,
which brings about nonvanishing $n_L$.
Fig.~\ref{fig:bubble_int} shows one of the dominant processes
that drives the CPV source terms, in this case the left-handed top density.
The Higgs bubble wall is denoted as the spacetime-dependent~\cite{xy} VEVs,
$v_{a}(x)$, $v_{b}(y)$ ($a,\,b=1,\,2$), and
the vertices are described by the interaction of Eq.~(\ref{Yukawa}).

With the closed time path formalism in the VEV insertion approximation,
the CPV source term $S_{ij}$ for left-handed fermion $f_{iL}$
induced by right-handed fermion $f_{jR}$ takes the form
\begin{align}
{S_{i_L j_R}(Z)}&= N_C F\,
    \text{Im}\big[(Y_1)_{ij}(Y_2)_{ij}^*\big]\, v^2(Z)\, \partial_{t_Z}\beta(Z),
\label{sourCPV}
\end{align}
where $Z = (t_Z,0,0,z)$ is the position in heat bath of very early Universe,
$N_C = 3$ is number of color, and $F$ is a function
(see Ref.~\cite{Chiang:2016vgf} for explicit form)
of complex energies of $f_{iL}$ and $f_{jR}$
that incorporate the $T$-dependent widths 
of particle/hole modes.
We note that, even though the angle $\beta$ is
basis-dependent in the general 2HDM,
its variation $\partial_{t_Z}\beta(Z)$ is physical~\cite{Del-beta} and
plays an essential role in generating the CPV source term.

If bubble wall expansion and $\partial_{t_Z}\beta(Z)$
reflect the departure from equilibrium,
the essence of the CPV for BAU is in $\text{Im}\big[(Y_1)_{ij}(Y_2)_{ij}^*\big]$.
Let us see how it depends on the couplings $\rho_{ij}$.
From Eqs.~(\ref{YD}) and (\ref{rhoij}), it follows that
\begin{align}
\text{Im}\big[(Y_1)_{ij}(Y_2)_{ij}^*\big]
= \text{Im}\big[(V_L^uY_{\rm diag} V_R^{u\dagger})_{ij}(V_L^u\rho V_R^{u\dagger})_{ij}^* \big].
\label{YCPV}
\end{align}
Suppose~\cite{Guo:2016ixx} $(Y_{1,\, 2})_{ij} = 0$, except for
$(Y_{1,\, 2})_{tc}$, and $(Y_1)_{tt}=(Y_2)_{tt}$,
with $t_\beta=1$ (which is maintained in this study) to simplify.
Then $\sqrt{2}Y^{\rm SM} = Y_1 + Y_2$ can be diagonalized by just $V_R^u$ to
a single nonvanishing $33$ element $y_t$, the SM Yukawa coupling,
while the combination $-Y_1 + Y_2$ is not diagonalized.
Solving for $V_R^u$ in terms of nonvanishing elements
in $Y_1$ and $Y_2$, one finds
\begin{align}
  \text{Im}\big[ (Y_1)_{tc}(Y_2)_{tc}^* \big]
 = -y_t\text{Im}(\rho_{tt}), \quad \rho_{ct}=0,
\label{simpCPV}
\end{align}
with $\rho_{tc}$ related to $\rho_{tt}$ but remaining a free parameter.
Although such a simple Yukawa texture makes it easy to see how
the BAU-related CPV emerges in the Yukawa sector at $T=0$,
the charm quark would be massless.
We therefore scan a wider parameter space, keeping the
physical Yukawa couplings in our numerical analysis.

To calculate $Y_B$, we need to calculate the density $n_L$ in Eq.~(\ref{YB}).
The relevant number densities are
$n_{q_3} = n_{t_L}+n_{b_L}$, $n_{t_R}$, $n_{c_R}$, $n_{b_R}$, and
$n_H=n_{H^+_1}+n_{H^0_1}+n_{H^+_2}+n_{H^0_2}$.
We solve a set of transport equations~\cite{Chung:2009qs}
that are diffusion equations fed by various density combinations
weighted by mass (hence $T$) dependent statistical factors,
but crucially also CPV source terms such as Eq.~(\ref{sourCPV}).

For our numerical estimates~\cite{errors},
we adopt the diffusion constants and thermal widths
of left- and right-handed fermions given in Ref.~\cite{DiffConst},
and follow Ref.~\cite{Huet:1995sh} to reduce the coupled equations
to a single equation for $n_H$,
controlled by a diffusion time $D_H \simeq 101.9/T$ modulated by $1/v_w^2$.
As discussed~\cite{EWPT_2HDM}, the EWPT has to be strongly first order.
In the current investigation, we use $T_C = 119.2$~GeV and $v_C = 176.7$ GeV,
which are calculated by using finite-temperature one-loop effective potential
with thermal resummation~\cite{Kanemura:2004ch},
taking $m_H = m_A = m_{H^\pm} = 500$ GeV,
$M \equiv m_3/\sqrt{s_\beta c_\beta} = 300$ GeV, and $t_\beta = 1$,
where $m_3$ is a mixing mass parameter between the two Higgs doublets $\Phi_{1,2}$.
In particular, we take $c_{\beta-\alpha} = 0.1$, which is close to alignment.
The chosen parameter set together with $\rho_{tt}$ specified below
are consistent with direct search bounds of the heavy Higgs bosons
at the LHC~\cite{LHC_heavyH}.
But the LHC should certainly have the ability to search for sub-TeV bosons.

\begin{figure}[t]
\center
\includegraphics[width=7cm]{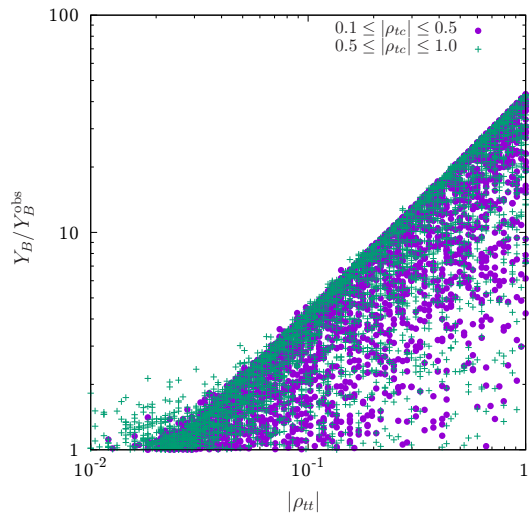}
\caption{Impact of $\rho_{tt}$ and $\rho_{tc}$ on $Y_B$,
where the phases $\phi_{tt}$ and $\phi_{tc}$ are scanned over 0 to $2\pi$,
with other parameters randomly chosen (see text for details).
The purple (green) points are for $0.1\le |\rho_{tc}|\le 0.5$
 ($0.5\le |\rho_{tc}|\le 1.0$).
}
\label{fig:Scat_YB_rhott}
\end{figure}

The $\rho_{ij}$s are
constrained~\cite{Crivellin:2012ye, Chen:2013qta, Altunkaynak:2015twa}
by $B_{d}$ and $B_{s}$ meson mixings and $b\to s\gamma$ decay.
In Ref.~\cite{Altunkaynak:2015twa}, it is found that
$|\rho_{tt}| < 2$, $|\rho_{tc}| < 1.5$ and $|\rho_{ct}| < 0.1$.
As a conservative choice, we consider
$|\rho_{tt}|$, $|\rho_{tc}| \le 1$ and $|\rho_{ct}| \le 0.1$,
with $\rho_{ij} = 0$ for all other entries.
Note that, from the observed flavor pattern and $y_t \simeq 1$,
having these two parameters at ${\cal O}(1)$ are the most reasonable.
Scanning over $\phi_{tt}$ and $\rho_{tc}$
 (but keeping a general texture such that physical 
 charm and top quark masses are kept),
we show $Y_B/Y_B^{\text{obs}}$ in Fig.~\ref{fig:Scat_YB_rhott}
as a function of $|\rho_{tt}|$.
The purple dots (green crosses) are for $0.1\le|\rho_{tc}|\le 0.5$
($0.5\le|\rho_{tc}|\le 1.0$),
and the phases $\phi_{tt}$ and $\phi_{tc}$ $\in (0, 2\pi)$.

We see that sufficient $Y_B$ can be generated over a large parameter space,
and that $|\rho_{tt}|$ is a stronger driver for $Y_B$ than $\rho_{tc}$,
as suggested by the simplified argument of Eq.~(\ref{simpCPV}).
However, for small $\rho_{tt} \lesssim 0.01$, large $\rho_{tc} = \mathcal{O}(1)$
with $|\sin\phi_{tc}| \simeq 1$ could come into play for EWBG.

\begin{figure}[t]
\center
\includegraphics[width=7.2cm]{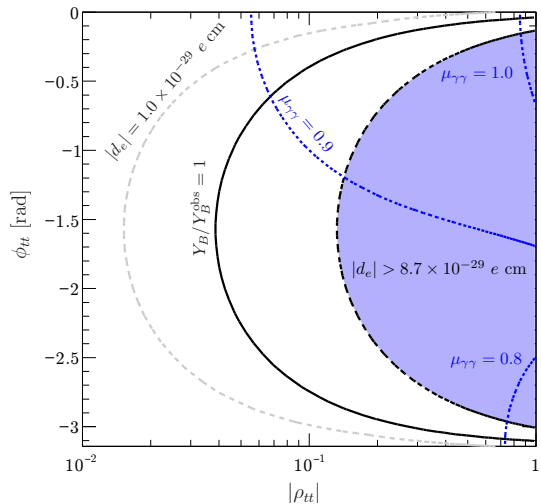}
\caption{
$Y_B$, $|d_e|$ and $\mu_{\gamma\gamma}$
on the $|\rho_{tt}|$--$\phi_{tt}$ plane, where
solid curve marks $Y_B/Y_B^{\text{obs}}=1$.
The shaded region
is excluded by the electron EDM bound,
with gray dashed curve its projected sensitivity.
The dotted curves are for $h \to \gamma\gamma$
with $\mu_{\gamma\gamma}$ as marked.
The $|d_e|$ and $\mu_{\gamma\gamma}$ results are for $c_{\beta-\alpha} = 0.1$.
}
\label{fig:BAUvsEDM}
\end{figure}

\paragraph{Phenomenological consequences.---}

Be it the $\rho_{tt}$ or $\rho_{tc}$-driven EWBG case,
a prominent signature would be $t\to ch$ decay~\cite{Hou:1991un}.
We find, for our benchmark, ${\cal B}(t\to ch)\simeq 0.15\%$
for $|\rho_{tc}|=1$ and $\rho_{ct} = 0$, which is
below the Run 1 bound of ${\cal B}(t\to ch) < 0.22\%$ ($0.40\%$)
from ATLAS~\cite{Aaboud:2017mfd} 
 (CMS~\cite{Khachatryan:2016atv}).
While search would continue at Run 2,
ATLAS has a projected reach~\cite{ATL-tch-gaga}
of ${\cal B}(t\to ch) < 0.015\%$ with full HL-LHC data,
based on $h \to \gamma\gamma$ mode alone.
Thus, the $\rho_{tc} \neq 0$ possibility is testable.
However, $t\to ch$ vanishes with $c_{\beta - \alpha} \to 0$,
 but a related signature for $\rho_{tc}\sim 1$ has been 
 studied~\cite{Kohda:2017fkn} recently. 
 The study shows that a search for $cg \to tH,~tA$ followed by 
 $H,~A\to t\bar{c}$ gives same-sign dilepton plus jets as signature, 
 which can be discovered with $300~{\rm fb}^{-1}$. 
 These complementary studies at the LHC would bring 
 powerful probes into the scenario.

Motivated by a hint~\cite{hmutau_LHC} for $h\to \mu\tau$ at LHC Run 1,
the case for nonzero $\rho_{\tau\tau}$ and$\rho_{\tau\mu}$
was explored for EWBG~\cite{Chiang:2016vgf}.
However, the recent CMS result~\cite{CMS:2017onh} of
$\mathcal{B}(h\to\mu\tau)<0.25\%$
has rendered this scenario unlikely.
For our current scenario, we find
${\cal B}(\tau\to\mu\gamma) \simeq (1\text{--}10)\times 10^{-9}$
for ${\cal B}(h\to \mu\tau) = 0.25~(0.1)\%$ with $0.05\lesssim |\rho_{tt}|\lesssim 0.5~(0.2\lesssim |\rho_{tt}|\lesssim 0.9)$ and $-\pi\lesssim \phi_{tt}\lesssim-0.8\pi$,
which can be probed by Belle II~\cite{Aushev:2010bq}
with sensitivity at $1\times 10^{-9}$.
Note that, due to destructive interference between one- and two-loop effects,
${\cal B}(\tau\to\mu\gamma)$ decreases as $\phi_{tt}\to 0$,
but ${\cal B}(h\to \mu\tau)$ and ${\cal B}(\tau\to\mu\gamma)$
vanish with $c_{\beta - \alpha} \to 0$.

A complex and sizable $\rho_{tt}$ can affect, through
the two-loop mechanism~\cite{Barr:1990vd}, the electron EDM,
where ACME has set a stringent limit~\cite{Baron:2013eja}
of $|d_e|<8.7\times 10^{-29}\ e\,{\rm cm}$ recently.
In Fig.~\ref{fig:BAUvsEDM}, the black solid curve marks
$Y_B/Y_B^{\rm obs}=1$ in the $|\rho_{tt}|$--$\phi_{tt}$ plane,
but the shaded region is excluded by the ACME bound,
which constrains $|\rho_{tt}|<0.1$--$0.2$ at $\phi_{tt} = -\pi/2$,
where there can still be sufficient BAU for $\rho_{tt}\gtrsim 0.04$.
The limit is expected~\cite{Baron:2016obh}
to improve down to $1.0\times 10^{-29}\ e\,{\rm cm}$ or better,
which is illustrated by the gray dashed curve.
Thus, electron EDM experiments probe the EWBG region in our scenario.

But the power of EDM probes brings about two issues. On one hand,
like previous cases, the $d_e$ constraint disappears with $c_{\beta - \alpha} \to 0$.
In addition, just like $\rho_{\mu\tau}$ and $\rho_{\tau\mu}$,
$\rho_{ij}$s in lepton sector need not vanish.
If $\rho_{ee}$ is turned on,
the value of $d_e$ could change considerably.
For $\rho_{ee} = y_e \equiv \sqrt{2}m_e/v$, the current $d_e$ bound
would exclude $|\rho_{tt}|\gtrsim 0.01$ for $\phi_{tt} = -\pi/2$.
However, for $\rho_{ee}=iy_e$,
cancellations could suppress $d_e$ in some region of
$|\rho_{tt}| \simeq 0.1$--1.0 and $-\pi\lesssim\phi_{tt}\lesssim -\pi/2$,
evading the current bound.
Even for such a case, however, the region that $d_e \simeq 0$
can be probed with the help~\cite{EDMcancel} of neutron and proton EDMs,
since the cancellation mechanism should not work simultaneously
for all EDMs.

Although our benchmark value of $c_{\beta-\alpha} = 0.1$
may seem small enough, the effects above
all vanish with $c_{\beta-\alpha} \to 0$, the alignment limit.
Other examples are, e.g. $A \to hZ$.
Alignment is quite effective in hiding the effects of the second Higgs doublet.
Are there effects that do not vanish with $c_{\beta - \alpha} \to 0$?
EWBG itself certainly is one.
Other important observables are $h\to \gamma\gamma$ decay
and $\lambda_{hhh}$ coupling, which are significantly modified
if the EWPT is strongly first order.

The charged Higgs $H^+$ would couple to $h$ and reduce the $h \to \gamma\gamma$ width,
while $\rho_{tt}$ affects the top loop, but would vanish with $c_{\beta - \alpha} \to 0$.
We illustrate our benchmark scenario with the blue dotted lines in Fig.~\ref{fig:BAUvsEDM}
for $\mu_{\gamma\gamma}=1.0$, 0.9 and 0.8 as marked.
In the alignment limit,
one has $\mu_{\gamma\gamma}\simeq 0.93$ from the charged Higgs boson loop,
where the actual number depends on Higgs sector details.
The combined Run 1 limit~\cite{Khachatryan:2016vau} from ATLAS and CMS
is $\mu_{\gamma\gamma}=1.14_{-0.18}^{+0.19}$.
Future measurements at the HL-LHC~\cite{HL-LHC},
ILC~\cite{Baer:2013cma} and CEPC~\cite{CEPC-SPPCStudyGroup:2015csa}
could improve the precision of $\mu_{\gamma\gamma}$ down to $\sim 5\%$,
hence $h \to \gamma\gamma$ would be an important test of the scenario.

The triple Higgs coupling $\lambda_{hhh}^{\text{2HDM}}$ receives
one-loop corrections~\cite{3h2HDM} that are proportional to
$m_\Phi^4[1-M^2/m_\Phi^2+ m_h^2/2m_\Phi^2]^3/v^3$,
where $\Phi=H^0, A^0, H^\pm$.
One sees that $\lambda_{hhh}^{\text{2HDM}}$ gets enhanced
by $m_\Phi^4$ if $m_{\Phi}$ receives substantial dynamical contributions
other than the common $M$.
We find $\Delta\lambda_{hhh}\equiv (\lambda_{hhh}^{\text{2HDM}}-\lambda_{hhh}^{\text{SM}})/\lambda_{hhh}^{\text{SM}}
\simeq 63\%$ for our benchmark point, taking subleading corrections into account.
Keeping Higgs sector parameters unchanged,
the number increases to $\simeq 74\%$ in the alignment limit.
 There are several prospects for measuring the triple Higgs coupling. 
 One is at the high luminosity LHC with $3000~{\rm fb}^{-1}$,  where 
 the accuracy amounts to $30-50\%$~\cite{Baglio:2012np,Goertz:2013kp, Barger:2013jfa}. Moreover, the International Linear Collider plans to measure the coupling 
 at $27\%$ accuracy with combined $250+500~$GeV data~\cite{Fujii:2017vwa}, 
 while future $100~$TeV colliders with $3000~{\rm fb}^{-1}$ can refine it to $8\%$ \cite{Yao:2013ika}. The future is wide open.

We have stated that our benchmark point,
in particular $m_{H^0} = m_{A^0} = m_{H^\pm} = 500$ GeV,
is not ruled out by LHC heavy Higgs search.
Our main purpose is to illustrate EWBG, and we have not made detailed
study of Higgs phenomenology, which would depend on the uncertain spectrum.
But ATLAS and CMS should reorient their $H^0$, $A^0$ and $H^\pm$ search
to the general 2HDM,
where phenomenology has been touched upon in Ref.~\cite{Hou:2017hiw}.
This reference has demonstrated that
alignment phenomenon emerges naturally in the general 2HDM
without $Z_2$ symmetry, with parameter space matching EWBG.

Finally, Ref.~\cite{Tulin:2011wi} used what we call $\rho_{ct}$
(but set $\rho_{tc} = 0$) to generate new CPV phases in $B_s$ mixing,
and suggested that the phase of $\rho_{tt}$ could drive EWBG,
but touched less on phenomenological consequences.

\paragraph{Conclusion.---}
We have studied EWBG induced by the top quark in the
general 2HDM with FCNH couplings.
The leading effect arises from the extra $\rho_{tt}$ Yukawa coupling,
where BAU can be in the right ballpark for $\rho_{tt}\gtrsim0.01$
with moderate CPV phase.
Even if $\rho_{tt} \ll 0.01$, $|\rho_{tc}|\simeq 1$ with large CPV phase
can still generate sufficient BAU.
These scenarios are testable in the future,
with new flavor parameters that have rich implications,
and extra Higgs bosons below the TeV scale.
Nature may opt for a second Higgs doublet for generating
the matter asymmetry of the Universe, through a new CPV phase associated with the top quark.


\vskip0.2cm
\noindent{\bf Acknowledgments} \
KF is supported in part by 
DOE contract DE-SC0011095,
ES is supported in part by
grant MOST 104-2811-M-008-011 and IBS under the project code, IBS-R018-D1, 
and WSH is supported by grants MOST 104-2112-M-002-017-MY2,
MOST 105-2112-M-002-018, NTU 106R8811 and NTU 106R104022.
WSH wishes to thank the hospitality of University of Edinburgh
for a pleasant visit.



\begin{thebibliography}{99}
\bibitem{xy}
  Spacetime dependence is due to wall thickness,
%
which is approximated by a hyperbolic tangent form.
\end{thebibliography}
\end{document}